# Electric Field in Materials, Containing Conductive Nanofibers

Yuri Kornyushin

*Maître Jean Brunschvig Research Unit,*
*Chalet Shalva, Randogne,*
*CH-3975, Switzerland*

Concentrated electric field and its energy in materials, containing nanofibers, are discussed. It is shown that the electric field in the vicinity of the end of a fiber is proportional to the external applied field and to the fiber length, whilst it is inversely proportional to the fiber diameter. Specific electrostatic energy of a fiber in a sample under the action of external applied field is calculated. This energy appears to be negative and proportional to the ratio of the fiber length to its diameter. This means that longer fibers are more stable than the shorter ones.

It is well known that materials, containing nanofibers possess many interesting and useful properties. Here we shall consider some of the electric properties of such materials. In [1] samples, containing parallel bundles of nanowires, submersed in a nonconductive media, were studied. External electric field, parallel to the bundles, had been applied. The high intensity of the field-induced electron emission and its low threshold were observed. These experimental data were considered by the authors to be the consequences of a small individual wire diameter and a long bundle length. The authors suggest that the electric field on the surface of a wire can be represented as $E = V/5r$ (here $V$ is the applied voltage, and $r$ is the radius of a wire). Here we suggest more detailed and specific explanation. In [2] the average value of the electric field in a nonconductive matrix, acting upon metallic fibers, $E_m$, is discussed. It is shown in [2] that the average value mentioned is described by the following relation:

$$E_m = \langle E \rangle /(1 - f). \qquad (1)$$

Here $\langle E \rangle$ denotes the uniformly applied in the direction of the fibers external field (its value is equal to that of the field averaged over the volume of a sample, $\langle E \rangle = V/L$, $L$ is the length of a sample), and $f$ is the ratio of the volume of all the inclusions (fibers) in a sample $v$ to the volume of a whole sample $V$. That is $f = v/V$.

Eq. (1) follows from the definition of the average field in a sample, and a well known fact, that an electric field in an isolated conductor is equal to zero, when the only external action on the conductor is an application of the external electric field. Really, the value of the field averaged over the whole volume of a sample is the sum of the average value of the field in inclusions $E_i$ multiplied by the volume of inclusions $v$ and the average value of the field in the matrix $E_m$ multiplied by the volume of a matrix $V - v$ divided by the volume of a whole sample $V$. That is $\langle E \rangle = [vE_i + (V - v)E_m]/V = fE_i + (1 - f)E_m$. As was pointed out above, the field inside the isolated conductive fiber $E_i = 0$. Taking this into account we get that from relation $\langle E \rangle = fE_i + (1 - f)E_m$ Eq. (1) follows.

At $f$ close to 1 (to the author best knowledge such samples were studied in [1]) the electric field, acting on the fibers (see Eq. (1)), is essentially larger than the external applied one.

There is something else. As in [1] the external field was applied along the fibers, the charges $\pm q$, giving rise to the electric field, which compensates that inside the fibers, occur only at the fiber ends, where a well known concentration of the electric field in the vicinity of sharp objects takes place [3]. As the length of the fiber $l$ in [1] is much larger than its diameter $d$, the capacity of both fiber ends $C$ could be approximated as $d/\pi$ [4]. In [4] a fiber with flat ends was studied. Electrostatic potentials of the field, produced by the charges $\pm q$, at the fiber ends could be estimated as $\varphi_{1,2} = \pm \pi q/\varepsilon d$ (here $\varepsilon$ is the dielectric constant of the matrix), and local electric field in the vicinity of the ends as $E_b = \pm 2\pi q/\varepsilon d^2$. Potential difference between the potentials on the two fiber ends could be approximated as $\delta\varphi = 2\pi q/\varepsilon d$. From the other hand the value of this potential difference is $lE_m$, so that the total potentials at both ends are equal. From this follows that $q = \varepsilon dlE_m/2\pi = \varepsilon dl\langle E\rangle/2\pi(1-f)$, and that the local concentrated field in the vicinity of the fiber ends is as follows:

$$E_b = \pm lE_m/d = \pm l\langle E\rangle/d(1-f). \qquad (2)$$

Here Eq. (1) was taken into account. When $f$ is comparable with unity, Eq. (2) is rather rough an approximation. It is worthwhile to note also that the field on the lateral surface of a conductive fiber is of a zero value and it increases with the increase of the distance from the fiber up to the distance equal to the fiber length.

These calculations show, that the electric field in the vicinity of the conductive fiber ends is $l/d(1-f)$ times larger than the external applied field $\langle E\rangle$, but average field in the matrix is only $1/(1-f)$ times larger than $\langle E\rangle$. Thus, the threshold of the field-induced electron emission is achieved already at a relatively low external applied field. In such a way the experimental data, obtained in [1] can be readily explained.

Charges at the fiber ends possess electrostatic energy. This energy consists of the intrinsic energy of these charges, $2\times 0{,}5q^2/\varepsilon C = \varepsilon(dl^2/4\pi)[\langle E\rangle/(1-f)]^2 = \varepsilon v_f(l/d)[\langle E\rangle/\pi(1-f)]^2$ (here $v_f = 0.25\pi d^2 l$ – the volume of one fiber), interaction energy of the charges with electrostatic field in the matrix, $-2\varepsilon v_f(l/d)[\langle E\rangle/\pi(1-f)]^2$, and electrostatic energy of the interaction of these two positive and negative charges situated on the fiber ends, $-\varepsilon(v_f/\pi)[\langle E\rangle/\pi(1-f)]^2$. Total electrostatic energy per unit volume of the fiber (that is the specific electrostatic energy) in the adopted in this paper approximation is as follows:

$$(U/v_f) = -\varepsilon[(l/d) + (1/\pi)][\langle E\rangle/\pi(1-f)]^2. \qquad (3)$$

It is worthwhile to note that specific electrostatic energy of the fiber contains a factor $\varepsilon(l/d)\langle E\rangle^2$. The value of the factor $\varepsilon(l/d)$ is essentially larger than unity.

In [5] electric conductivity of PVC films was studied. Spontaneous reversible transitions between the two states of the films, with high and low conductivity, were detected. It is reported in [5] that studied PVC films have a complex micro- and macromolecular structure and contain microscopic quasicrystalline and amorphous domains. Such a complex structure of the films allows authors to suggest (with a view to understanding the experimental results, obtained by the authors) the presence of domains with relatively high conductivity.

It is worthwhile to note that negative specific electrostatic energy of domains discussed in the present paper, as one can see from Eq. (3), contributes to the domain stability at its elongation, as its specific electrostatic energy is decreased essentially during the elongation. When performing a detailed thermodynamic analysis of regarded system, the positive surface energy of the surface of fibers in a non-conductive matrix should be taken into account.

Conductivity jumps reported in [5] could occur as a result of growing of some number of high conductivity fibers through the film, or overlapping of the fibers, leading to the percolation process. Decrease in the electrostatic energy in the course of these processes contributes to their advance. One cannot also exclude the contribution of the field-induced electron emission to the observed increase in the electric conductivity of the samples, the same way as it was also found in [1].

Observed and reported in [5] transitions could be a manifestation of some first-order phase transformation. In this case one can expect that the direct transition takes place at a greater value of the external applied field compared with the value of the field at which the reverse transition occurs. Unfortunately, the phenomenon of the presence or absence of hysteresis mentioned here is not discussed in [5].